\documentclass[fleqn,usenatbib]{mnras}
\def\paren#1{\left( #1 \right)}
\def\bra#1{\left[ #1 \right]}
\def\angl#1{\left\langle #1 \right\rangle}
\def\ltsima{$\; \buildrel < \over \sim \;$}
\def\lsim{\lower.5ex\hbox{\ltsima}}
\def\gtsima{$\; \buildrel > \over \sim \;$}
\def\gsim{\lower.5ex\hbox{\gtsima}}
\newcommand{\bhat}[1]{{\hat {\bf #1}}}
\usepackage[T1]{fontenc}
\usepackage{ae,aecompl}
\usepackage{graphicx}
\usepackage{epstopdf}
\usepackage{amsmath}
\usepackage{amssymb}
\usepackage{enumerate}

\title[Tidal Disruption of Inclined or Eccentric Binaries by Massive Black Holes]
{Tidal Disruption of Inclined or Eccentric Binaries by Massive Black Holes}
\author[Brown, H., Kobayashi, S., Rossi E. M. and Sari, R]
{Harriet Brown$^{1}$, Shiho Kobayashi$^{1}$, Elena M. Rossi$^2$ and Re'em Sari$^3$
\\
$^{1}$Astrophysics Research Institute, LJMU, IC2, Liverpool Science Park, 
146 Brownlow Hill, Liverpool L3 5RF, UK
\\
$^{2}$Leiden University, Oort Building, Niels Bohrweg 2, 2333 CA Leiden, The Netherlands
\\
$^{3}$Racah Institute of Physics, Hebrew University, Jerusalem 91904, Israel
}
\date{Accepted XXX. Received YYY; in original form ZZZ}
\pubyear{2017}
\begin{document}
\label{firstpage}
\pagerange{\pageref{firstpage}--\pageref{lastpage}}
\maketitle

\begin{abstract}
Binary stars that are on close orbits around massive black holes
(MBH) such as Sgr A* in the centre of the Milky Way are liable to undergo
tidal disruption and eject a hypervelocity star. 
We study the interaction between such a MBH and circular binaries 
for general binary orientations and penetration depths (i.e. binaries 
penetrate into the tidal radius around the BH). We show that for very deep 
penetrators, almost all binaries are disrupted 
when the binary rotation axis is roughly oriented toward the BH 
or it is in the opposite direction. 
The surviving chance becomes significant when the angle between the binary rotation 
axis and the BH direction is 
between $0.15\pi$ and $0.85\pi$.
The surviving chance is as high as 
$\sim 20$\% when the binary rotation axis is perpendicular to the BH direction. 
However, for shallow penetrators, the highest disruption chance is found 
in such a perpendicular case, especially in the prograde case. This is because 
the dynamics of shallow penetrators is more sensitive to the relative orientation of the binary and orbital
angular momenta.
We provide numerical fits to the disruption probability and energy gain at the 
the BH encounter as a function of the penetration depth. The latter can be simply 
rescaled in terms of binary masses, their initial separation and the binary-to-BH 
mass ratio to evaluate the ejection velocity of a binary members in various systems.
We also investigate the disruption of coplanar, eccentric binaries
by a MBH. It is shown that for highly eccentric binaries retrograde orbits
have a significantly increased disruption probability and ejection velocities 
compared to the circular binaries.
\end{abstract}
\section{Introduction}
Hypervelocity stars (HVSs) are stars with sufficient velocity to escape from the
Galactic gravitational potential. Targeted HVS Surveys 
\citep{2009ApJ...690.1639B,2012ApJ...751...55B,2014ApJ...787...89B}
have lead to the identification of 21 unbound stars to date. 
There are two main processes theorized to produce
HVSs from the Galactic nucleus: 
the disruption of a binary system by a massive black hole 
(MBH) know as the Hills mechanism \citep{1988Natur.331..687H}, and 
three-body interaction between a MBH binary and 
an orbiting star \citep{2003ApJ...599.1129Y}. 
For a binary with separation $a$ and total mass $m$ interacting with a MBH
with mass $M$, the distance at which the tidal forces overcome the binary's self-gravity 
is about $r_t = a(M/m)^{1/3}$.  According to the Hills mechanism, 
once the binary crosses the tidal radius and it is tidally disrupted, 
one of the binary members is ejected at high speeds of the order of 
$v_0 (M/m)^{1/6} \sim 2000 (m/M_\odot)^{1/3}(a/5R_\odot)^{-1/2}(M/10^6M_\odot)^{1/6}$ km/s
where $M_\odot$ and $R_\odot$ are the solar mass and radius, respectively
\footnote{Velocities in the Galactic halo are lower due to the Galactic potential.}. 
The other binary member is bound to the MBH.  The Galactic centre hosts a population of 
young, massive stars which have eccentric, randomly distributed orbits 
(e.g. \citealt{2008ApJ...689.1044G,2009ApJ...692.1075G}). These S-stars are considered
to be the counterparts of the HVSs. 
There has been significant theoretical work 
that has gone into modelling the results of binary tidal disruption events 
(e.g. \citealt{2005MNRAS.363..223G,2006MNRAS.368..221G,2006ApJ...653.1194B,
2007MNRAS.379L..45S,2010ApJ...708..605S,2010ApJ...709.1356L,2010ApJ...713...90A,
2012ApJ...748..105K}).

Previous numerical studies by \cite{2006ApJ...653.1194B} have
found the disruption probability of a binary at the encounter with a MBH is 
roughly linear with its penetration depth (the ratio of 
the closest approach distance
to it's tidal radius). However, these simulations do not fully explore 
the deepest penetrations and utilize a full 
three-body model which is relatively computationally expensive, limiting the parameter
space that one would be able to reasonably explore. In order to efficiently 
explore the parameter space, a restricted three-body approximation was proposed 
\citep{2010ApJ...708..605S,2012ApJ...748..105K}, and it has been shown that the approximation 
is very accurate when the binary-to-BH mass ratio is large 
$M/m \gg 1$. In this method, the essential system parameters are only the 
binary orientation, the binary phase and the penetration depth, and we can obtain 
analytic solutions when binaries deeply penetrate the BH tidal radius. 
This method has also been used to 
model the velocity distribution of HVSs \citep{2014ApJ...795..125R}, and to fit for the 
first time current data to given a constrain on the binary properties and the 
Galactic Potential \citep{2017MNRAS.467.1844R}.

All work done utilizing this method has so far only examined circular binaries 
that are co-planar with their orbit around a MBH. 
In this paper, we will apply the method to non-coplanar binaries, and we examine 
how the binary orientation affects the disruption probability and ejection velocities. 
In \S2 we describe the restricted three-body approximation, and we discuss how 
symmetry in the system can be used to further reduce the volume of the parameter space.
In \S3 we numerical obtain the disruption rate of binaries and the energy gain at 
the BH encounter, and we compare with previous theoretical models.  
We also discuss the fate of coplanar, eccentric binaries. 
The conclusions and the implications of our results are discussed in \S4.

\section{Parabolic and Radial Restricted Three-Body Approximations}
In order to discuss the tidal encounter of binaries with a MBH,
we employ the restricted 3-body approximation presented by \cite{2010ApJ...708..605S}, which is valid when the binary mass is much smaller than that of the MBH. In the following discussion, we assume that the masses of the two binary members, the primary $m_1$ and the secondary $m_2$ (the total mass $m=m_1+m_2$), are of the order of solar mass, and the MBH mass $M$ is similar to that of the MBH at the Galactic centre. Although the exact values of the masses are not important in our formulation, the large mass ratio $M/m\gg1$ ensures our approximation.

In this approximation,  the relative motion of the two binary members 
$\bf r= r_2-r_1$ can be formulated as the motion of a single particle under 
the influence of external time-dependent forces. 
We apply this approximation to a binary system 
injected in a parabolic orbit ${\bf r}_m$ with
periapsis $r_p$ around a MBH. 
Rescaling the distance between the binary members by $(m/M)^{1/3} r_p$ and 
the time by $\sqrt{r_p^3/GM}$,
the equation of motion is given in terms of the dimensionless variables 
${\boldsymbol \eta}\equiv  (M/m)^{1/3}  ({\bf r}/r_{\rm p})$ and $t$ as
\begin{equation}
\ddot {\boldsymbol \eta}=\left( r_p  \over r_{\rm m}\right)^3
\left[ - {\boldsymbol \eta}+3 ({\boldsymbol \eta} \cdot \bhat r_{\rm m}) 
\bhat r_{\rm m} \right]
- { {\boldsymbol \eta} \over |{\boldsymbol \eta}|^3},
\label{eq:r_vec}
\end{equation}
where $\bhat r_{\rm m}={\bf r}_m/r_m=(\cos f, \sin f, 0)$ is 
a unit vector pointing the centre of mass of the binary, 
the distance from the MBH is given by 
$r_{\rm m}=2r_{\rm p}/(1+\cos f)$,
and $f$ is the angle from the point of closest approach. 
The angle $f$, known as the 
true anomaly, is a function of 
time, but analytically one has only the time as a function of $f$. For numerical 
applications it is preferable to use its differential (and dimensionless) form $\dot{f}=\sqrt{2}(1+\cos f)^2/4$. In Section 3, we will numerically integrate Equation (1) and this $\dot{f}$ equation together.
We start at a radius well outside the tidal sphere $r_m \gg r_t$, and
we evaluate the entire evolution of the binary system by using these equations.

If a binary is ejected toward a MBH around the radius 
of influence of the BH, the orbital energy is negligible compared to the 
the energy gain or loss of each binary member at the BH encounter. 
As we have shown, parabolic orbits can be used 
for the binary's centre of mass to evaluate the characteristic of HVSs 
\citep{2012ApJ...748..105K}. 
Since the self-gravity energy of the binary is smaller by a factor 
of $(M/m)^{1/3}\gg 1$ than the energy gain or loss at the BH encounter,
it can also be neglected. This means that the total energy of 
the system is zero, and the energies of the primary and secondary members 
are related as $E_1=-E_2$. In terms of our dimensionless 
Cartesian coordinates ${\boldsymbol \eta}=(x,y,z)$,
the energy is given \citep{2010ApJ...708..605S,2012ApJ...748..105K} by 
\begin{equation}
\label{E1par}
\begin{split} 
E_2  = {G m_1m_2 \over a\, D} \left(M\over m\right)^{1/3} 
\left[ {\left(1+\cos f\right)^2 \over 4}(x\cos f+y\sin f) \right. \\ 
\left. + \frac{-\sin f \dot x +(1+\cos f) \dot y }{\sqrt 2} \right] , 
\end{split}
\end{equation}
where $D=r_p/r_t$ is the penetration factor and $r_t$ is the tidal radius. $D$ indicates how deeply the binary penetrates into the tidal sphere.
If the binary dissolves at the BH encounter, this energy becomes a constant, 
since each binary member is eventually moving only under the conservative force 
of the BH.

The angular momentum of each binary member around the MBH,
${\bf L}_i=m_i ({\bf r}_i \times \dot{\bf r}_i )$ 
($i=$1 or 2), also becomes a constant of the motion 
when the binary is disrupted. Considering that the positions of the members can be expressed by using the centre of mass ${\bf r}_m$ and the displacement vector
${\bf r}$ as ${\bf r}_1={\bf r}_m-(m_2/m){\bf r}$ and 
${\bf r}_2={\bf r}_m+(m_1/m){\bf r}$, 
we can rewrite the angular momenta up to the linear order of $r$ as 
\begin{equation}
 {\bf L}_i = \paren{\frac{m_i}{m}}{\bf L}_m +\Delta {\bf L}_i, 
\label{eq:angmom}
\end{equation}
where ${\bf L}_m$ is the angular momentum of the centre of mass and 
$\Delta {\bf L}_i$ is the angular momentum change of the members at the tidal encounter,
\begin{eqnarray}
{\bf L}_m &\equiv& m ({\bf r}_m \times \dot{\bf r}_m) = \paren{0,0, \sqrt{2GMm^2r_p}}, \\
\Delta {\bf L}_2 &=& -\Delta {\bf L}_1 \equiv
 \paren{\frac{m_1m_2}{m}}
\bra{{\bf r}_m\times \dot{\bf r}+ {\bf r}\times \dot{\bf r}_m}.
\end{eqnarray}
$\Delta {\bf L}_i$ become constant vectors when the binary is disrupted. 
The exact values can be estimated by using Equation (\ref{eq:r_vec}) 
in a similar way as we have evaluated $E_i$. However, 
considering $r_m \sim r_t$ and $|\dot{\bf r}_m|\sim \sqrt{GM/r_t}$ 
at the binary disruption, $|\Delta {\bf L}_i|$ is about
$(M/m)^{1/3} \sqrt{Gm^3a}$. This is smaller by a factor of 
$\sim D^{1/2} (M/m)^{1/3}\gg 1$ than $L_m$. The angular momentum 
change at the tidal encounter is not important if it is not a very 
deep encounter. 

After the tidal disruption, the eccentricities of their orbits around the MBH
are given by 
$e_i =\sqrt{1+2L_i^2 E_i/m_i^3G^2M^2}$. If $E_i<0$ ($E_i>0$), 
the member is captured (ejected). 
For $D \gg (m/M)^{2/3}$, we can neglect the linear terms $\Delta {\bf L}_i$ 
in their angular momenta, and we obtain the eccentricity of the captured member as 
\begin{equation}
1-e_i \sim 2D\paren{\frac{m_1m_2}{m_i m}}\paren{\frac{m}{M}}^{1/3}  |\bar{E}|.
\label{eq:ecc}
\end{equation}
where $\bar{E}$ is the energy gain $E_2$ in units of $(Gm_1m_2/a)(M/m)^{1/3}$. Since $|\bar{E}|$ is expected to be of order of unity, 
the bound orbit would be very eccentric \citep{1991AJ....102..704H,2005ApJ...626..849P,2012ApJ...748..105K}.

\begin{figure}
\includegraphics[width=\columnwidth]{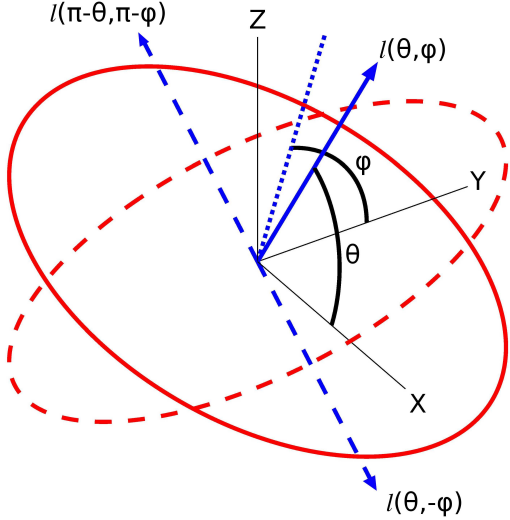}
\caption{The initial circular orbit of a binary (red solid line) and 
its angular momentum vector (blue solid arrow) with its projection (blue dotted line) 
on the Y-Z plane. The binary itself orbits the MBH on a parabolic orbit on the X-Y plane. 
The red dashed line indicates another circular orbit which is symmetric to the 
red solid orbit with respect to the X-Y plane. The two blue dashed arrows are the 
angular momenta of the same (red dashed) orbit but orbiting in the opposite direction.}
\label{fig:inital_setup}
\end{figure}

Since in this framework, results can be simply rescaled in terms of binary masses,
their initial separation, and the binary-to-BH mass ratio, the system is essentially 
characterized by four parameters for circular binaries: 
the penetration factor $D$, the initial binary phase $\phi$ 
and the orientation $(\theta, \varphi)$ of the binary's angular momentum
where $\theta$ is the inclination angle measured from the positive $x$-axis, and $\varphi$ is 
the azimuth angle measured from the positive $y$-axis on the $y$-$z$ plane 
(see the blue solid arrow in Figure \ref{fig:inital_setup}). 
For eccentric binaries, we have two additional parameters: the eccentricity $e$ of the orbits and 
the direction of the semi-major axis. We characterize the latter by the vector connecting 
the binary's centre of mass to the secondary member's periapsis. Although our formalism 
can be applied to explore the fate of a binary with an arbitrary orbit orientation, we will 
discuss only coplanar binaries when we numerically study the evolution of eccentric binaries 
in the next section. The direction of the vector (i.e. the direction of the semi-major axis) 
will be given by an angle $\varpi$ measured from the positive $x$-axis.

By considering two kinds of pairing of solutions which originate from symmetry in the system,
we can further reduce the volume of the parameter space. 
\begin{itemize}
 \item The negative of a solution is also a solution 
for eq. (\ref{eq:r_vec}). However, since the energy eq. (\ref{E1par}) is also linear in the coordinates, for circular binaries, a body starting 
with a phase difference $\pi$ will have the same final energy in absolute value but 
the opposite in sign \citep{2010ApJ...708..605S}. The ejected (captured) member is captured 
(ejected) if the initial binary phase is increased by $\pi$.
We just need to sample the binary phase $\phi$ between 0 and $\pi$.
For eccentric binaries, this is translated to the orientation of the semi-major axis, 
$\varpi$ should be sampled between 0 and $\pi$ (the binary phase $\phi$ should be 
considered between 0 and $2\pi$). For non-coplanar eccentric binaries, $\varpi$ should be redefined 
appropriately (e.g. if the direction vector of the semi-major axis is projected in
the $x$-$y$ plane, we would use the angle between the projected vector and x-axis), 
but we still need to sample it only between 0 and $\pi$.

\item Another kind of pairing is possible if one notices that the system is symmetric 
with respect to the $x$-$y$ plane. If $\{x(t), y(t),z(t)\}$ is a solution, 
$\{x(t), y(t), -z(t)\}$ is also a solution (see the red solid and dashed lines in 
Figure \ref{fig:inital_setup}). Since the energy eq. (\ref{E1par})  
does not depend on $z(t)$, they have the same energy as expected.
If the orientation of a binary is parametrised by ($\theta$, $\varphi$) as 
$\hat{l}=(\cos \theta, \sin \theta \cos \varphi, \sin \theta \sin \varphi)$, 
a sphere is defined by $\hat{l}$ for generic ($\theta$, $\varphi$). 
One might think at first that there is correspondence between points on the sphere 
which are located symmetrically with respects to the $x$-$y$ plane: 
$(\theta, \varphi) \leftrightarrow (\theta, -\varphi)$. However, the binaries should 
rotate in the same direction when they are projected in the $x$-$y$ plane. 
The correspondence actually exists between points symmetric about $z$-axis: 
$(\theta, \varphi) \leftrightarrow (\pi-\theta, \pi-\varphi)$. Note that 
$(\theta, -\varphi)$ and $(\pi-\theta, \pi-\varphi)$ indicate the same binary 
orientation except the rotation direction (i.e. clockwise or anticlockwise).
To investigate how the orientation of circular or eccentric binaries affects 
the disruption process, it is sufficient if we consider only the hemisphere 
defined by $0\le \theta \le \pi/2$ and $0 \le \varphi<2\pi$ (i.e. the fore-side of the sphere). 
\end{itemize}

\subsection{Radial approximation} 
In the limit of deep penetrations 
$D \ll 1$, the trajectory of the binary's centre of mass becomes 
almost radial. By assuming a radial orbit for the trajectory, we can obtain another 
set of approximation formulae. 
This radial approximation is useful when we investigate the binary disruption process 
in the deep penetration limit.
Since the binary orientation is determined by a single parameter (i.e. 
the angle between the binary rotation axis and the radial direction), 
the discussion is simpler. However, the difficulty arises from 
the assumption of a purely radial orbit with which the binary goes straight towards the BH.
A deep parabolic orbit with $D\ll 1$ parallels closely the radial one and gets around 
the BH smoothly. Since the energy gained by one of the binary members (or the energy 
loss by the other) during the periapsis passage is smaller by a factor of $D^2 \ll 1$ 
than gained (or lost) around the tidal radius, the perturbations caused by the binary 
mutual gravity is negligible around the passage, each of the binary members turns around 
with a constant orbital energy (see \cite{2010ApJ...708..605S} for the details). 
Therefore, we can connect an incoming radial orbit
with an outgoing radial one. For this purpose, we use free solutions which are available 
when the binary is well within the tidal radius.  Since around the periapsis 
passage ($t=0$), the BH tides dominate over the mutual gravity of the binary,
the last term in the right-hand side of Eq (\ref{eq:r_vec}) is negligible.
When the centre of mass of the binary moves on a radial orbit, 
the relative position of the binary members $\mathbf{r}=\mathbf{r}_2-\mathbf{r}_1$ are given 
by \citep{2010ApJ...708..605S},
\begin{eqnarray}
\label{freerad}
x(t) & = & A_x |t|^{-1/3}+ B_x |t|^{4/3}, \nonumber \\ 
y(t) & = & A_y |t|^{1/3}+ B_y  |t|^{2/3},\\
z(t) & = & A_z |t|^{1/3}+ B_z  |t|^{2/3}, \nonumber
\end{eqnarray}
where the distances and the time have been scaled by the initial semi-major axis of the binary 
$a$ and the inverse angular frequency of the binary $\sqrt{Gm/a^3}$, respectively.
We can ferry the free solution across $t=0$, from negative to positive times by simply 
changing the sign of the $A$ coefficients.
More specifically, we evaluate the six coefficients by using 
the numerical position $(x,y,z)$ and velocity $(v_x,v_y,v_z)$ 
at $t=-t_{min} <0$. Then, we resume the radial three-body approximation calculations 
at $t= t_{min} >0$.  We have evaluated the evolution of the system based on 
the radial approximation with different values of $t_{min}$, and we find that the 
difference in the ejection energy becomes less than 0.1\% for $t_{min} < 10^{-6}$.  
We will use $t_{min} = 10^{-6}$ for radial approximation calculations. 
\section{Numerical Results}

\subsection{Circular Binaries}
Our numerical calculations utilize a 4th-order Runge-Kutta method to integrate
the equations of motion. The time steps of the integration are scaled
by the minimum value of the three dynamical times associated with the binary pair and 
the interaction between each binary member and the BH. Circular binaries are
injected in parabolic orbits around a MBH.
To uniformly sample the binary orientation, we populate the surface of a unit sphere 
with equally spaced 2000 grid points. The regular equidistribution can be achieved 
by choosing circles of latitude at constant intervals $d_\theta$ and on these 
circles points with distance $d_\varphi \sim d_\theta$. 
For each grid point, the binary phase $\phi$ is sampled with 200 equally spaced 
grid points between 0 and $\pi$. 

\begin{figure}
\includegraphics[width=\columnwidth]{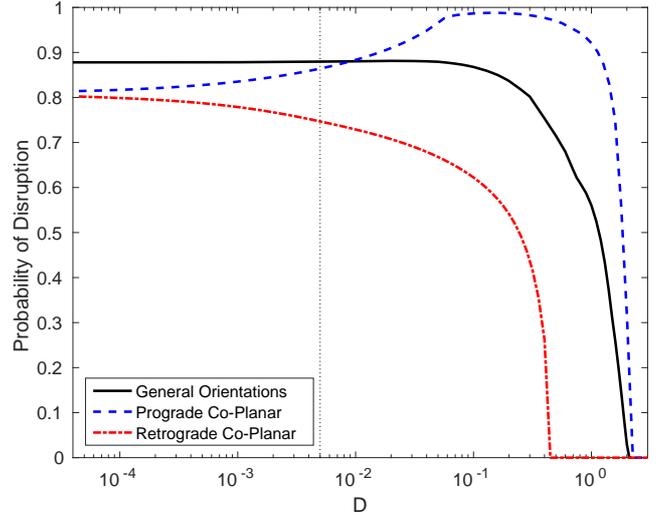}
\caption{Probability of disruption as a function of the penetration factor $D$. General binary orientations (black solid), coplanar prograde orbits (blue dashed), and coplanar retrograde orbits (red dot-dashed). 
The vertical dotted line marks the penetration limit for 
binaries of solar-type stars with $a=1$AU where the individual stars undergo tidal disruption.}
\label{fig:dis_with_D}
\end{figure}

Figure \ref{fig:dis_with_D} indicates the probability of binary disruption at 
the BH encounter as a function of $D$ averaged over phase and orientation. 
The largest $D$ for which there is disruption is $D=2.1$ for 
the coplanar prograde orbits, and for all sampled orbits. This indicates that 
coplanar prograde orbits have the highest disruption chance for the shallow 
encounters. For shallow penetrators $D \sim 1$-$2$, the disruption
probability is approximately linear with D.  
\cite{2006ApJ...653.1194B} have reported a linear relationship 
$P_{dis}\sim 1-D/2.2$. However, for smaller $D$, the
disruption rate plateaus with $\sim  88\%$ (the black solid line). Interestingly 
about $12 \%$ of binaries survive the BH encounter even for very 
deep penetrators $D\ll 1$. 
Our numerical results can be well approximated by a 5th-order polynomial,
\begin{equation}
P_{dis}(D) = A_0 + A_1D + A_2D^2 + A_3D^3 + A_4D^4 + A_5D^5,
\end{equation}
with coefficients: $A_0=0.8830$, $A_1=-0.0809$, $A_2=-1.0541$, 
$A_3=1.5377$, $A_4=-0.9249$, $A_5=0.1881$, for $D < 2.1$. 
The fractional error $\Delta P_{dis}/P_{dis}$ is less than $1\%$ for $D \la 1$. 
As the disruption probability approaches zero around $D\sim 2$, the fractional error 
becomes larger, but it is still about $5\%$ at $D=1.8$ and about $20\%$ at $D=2$.
This disruption rate ${88}\%$ at $D\ll 1$ is higher than that for coplanar binaries. 
Both coplanar prograde (the blue dot-dashed line) and retrograde 
(the red dot-dashed line) cases saturate at a level of $\sim 80\%$ 
\citep{2010ApJ...708..605S}. 

The disruption rate estimates for $D\ll 1$ break down when the pericentre 
distance to the MBH 
becomes comparable to the tidal disruption radius of 
the binary members (i.e. individual stars). If the binary members are solar-type stars 
with radius $R_\odot$ and its initial separation is $a=1$ AU, the stars themselves 
are disrupted for $D\lsim R_\odot/a \sim 5\times 10^{-3}$ (the vertical dotted line 
in Figure \ref{fig:dis_with_D}). To achieve a smaller $D$, the initial separation $a$ should be  wider, 
or the binary members should be compact objects such as stellar mass BHs, 
neutron stars and white dwarfs. 
The evolution of a stellar mass BH binary should be well described by the 
point-particle model. However, if the periapsis is
close to the event horizon scale $R_g$ of the central MBH, the Newtonian 
formulation would break down. Relativistic effects are negligible for 
$D \gg (m/M)^{1/3}R_g/a \sim 8\times 10^{-4} (a/1AU)^{-1} (m/4M_\odot)^{1/3}
(M/4\times10^6M_\odot)^{2/3}$.

\begin{figure}
\includegraphics[width=\columnwidth]{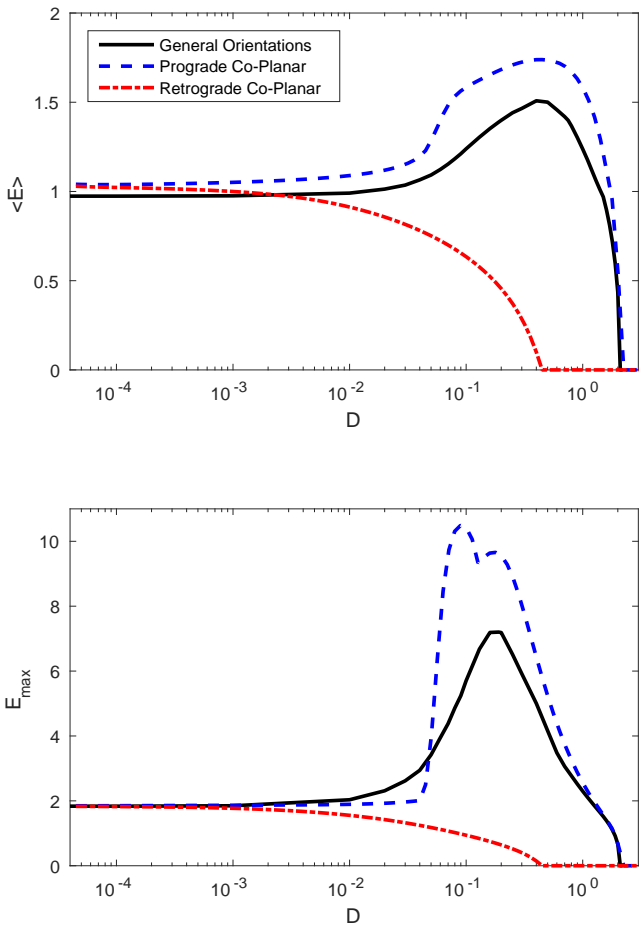}
\caption{Top panel: Ejection energy averaged over binary phase and orientation as a function of $D$.
Bottom panel: Characteristic maximum ejection energy as a function of $D$.
For a given $D$, the top $1\%$ have ejection energy higher than the characteristic maximum energy $E_{max}$.
General binary orientations (black solid), coplanar prograde orbits (blue dashed), 
and coplanar retrograde orbits (red dot-dashed). 
The average and characteristic maximum energy 
are in units of $(Gm_1 m_2 /a) (M/m)^{1/3}$, and they are evaluated for the absolute value of 
the energy $|E|$.}
\label{fig:energy_with_D}
\end{figure}

The top panel of Figure \ref{fig:energy_with_D} shows the ejection energy averaged over the binary phase and 
orientation as a function of $D$. We also plot in the bottom panel the characteristic 
 maximum ejection energy $E_{max}$ for a given $D$ as a function of $D$. 
This is estimated to characterize the population of the highest energy gain cases,
the top $1\%$ of the sampled cases have ejection energy higher than this energy. 
This threshold value is rather insensitive to the grid 
resolution, compared to the actual maximum value  which is as high as 
$\sim 27$ for a coplanar prograde orbit with $D\sim 0.1$. In both plots, a peak 
is present (the black solid lines), and the peak values are lower than for the prograde 
orbits (the blue dot-dashed lines). There are two 
peaks in the $E_{max}$ distribution for the prograde orbits. 
The average energy is approximated by a polynomial,
\begin{equation}
\begin{split}
\langle E \rangle = A_0 + A_1D + A_2D^2 + A_3D^3 + A_4D^4 + A_5D^5,
\end{split}
\end{equation}
with coefficients: $A_0=0.9582$, $A_1=3.3268$, $A_2=-6.6801$, $A_3=5.2785$, $A_4=-1.8731$, $A_5=0.2260$, where this energy is in units of $(Gm_1 m_2 /a) (M/m)^{1/3}$.
The fractional error is less than $1\%$ for $D \la 1$, and it is about 
$3\%$ at $D=1.8$ and about $10\%$ at $D=2$. By equalizing this energy in the physical units 
with the kinetic energy $m_1v_1^2/2$ (or $m_2v_2^2/2$), we can estimate the ejection velocity 
of the primary (or secondary) star at a distant place from the BH. The Galactic potential 
should be taken into account separately to estimate the velocity in the halo 
(e.g. \citealt{2014ApJ...795..125R,2017MNRAS.467.1844R}).

\begin{figure}
\includegraphics[width=\columnwidth]{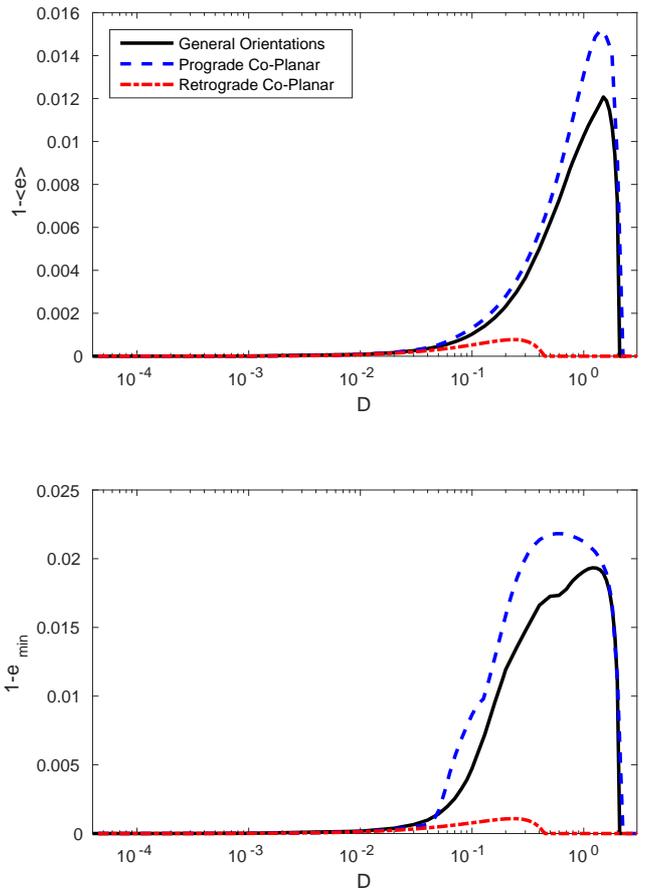}
\caption{Top panel: Mean eccentricity difference from a parabolic orbit in bound stars as a function of $D$. Bottom panel: Maximum eccentricity difference from a parabolic orbit in bound stars as a function of $D$, the bottom $1\%$ have eccentricity lower than the 
characteristic minimum eccentricity $e_{min}$.
General binary orientation (black solid), coplanar prograde orbits (blue dashed), 
and coplanar retrograde orbits (red dot-dashed). $m_1=m_2$ and $M/m=10^6$ are assumed.}
\label{fig:e_comp}
\end{figure}

The eccentricities of bound stars are given by 
$1-e \sim D(m/M)^{1/3}|\bar{E}|$ for equal mass 
binaries. Assuming $M/m=10^6$, the mean eccentricity 
difference $1-\angl{e}$ and the characteristic maximum 
difference $1-e_{min}$ are shown as a function of $D$
in Figure \ref{fig:e_comp}. If we consider deep penetrators, 
since $\angl{\bar{E}} \sim 1$ and $E_{max} \sim 2$ 
for $D\ll 1$ (see Figure \ref{fig:energy_with_D}), 
the mean value is given by 
$1-\angl{e} \sim 10^{-2} D$  and 
$1-e_{min}$ is larger by a factor of $\sim 2$.
For very deep penetrators 
$D\lsim (m/M)^{2/3} =10^{-4}$, the distributions 
flatten out even in a log-log plot
as $\Delta {\bf L}_i$ contributes to ${\bf L}_i$
(since the behaviour around $D\sim 1$ is more 
important in the context of HVS study, $1-e$ is plotted in 
the linear scale). 
Shallow penetrators $D\sim 1$ give lower eccentricities, 
but they are still very high $e\sim 0.98-0.99$ 
\citep{2005ApJ...631..L117,2007ApJ...656..709}
for S-stars in the Galactic centre ($0.3 \lsim e \lsim 0.95$; \cite{2009ApJ...692.1075G,2008ApJ...689.1044G}), suggesting that post-capture relaxation is the significant factor in determining S-star eccentricities \citep{2009ApJ...702..884P,2017ARAA.55.1}.

\begin{figure}
\includegraphics[width=\columnwidth]{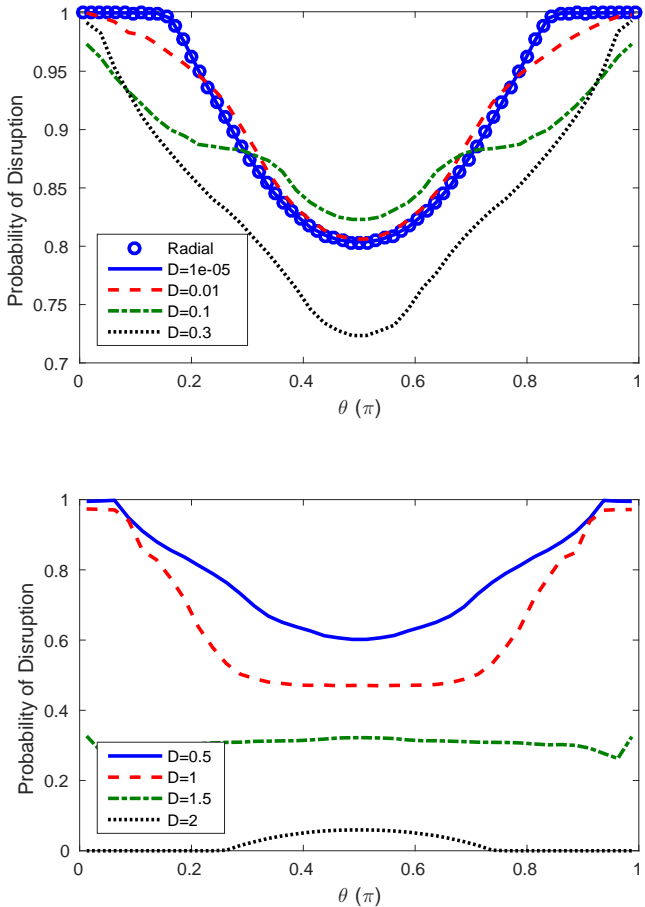}
\caption{Probability of disruption for a given $D$ as a function of the inclination angle 
$\theta$. Upper panel: deeper penetrators: the radial approximation (blue circles) and 
the parabolic approximation with $D=10^{-5}$ (blue solid), $10^{-2}$ (red dashed), $10^{-1}$ (green dot-dashed), and $0.3$ (black dotted). Bottom panel: shallower penetrators: 
the parabolic approximation with $D=0.5$ (blue solid), $1.0$ (red dashed), $1.5$ 
(green dot-dashed) and $2.0$ (black dotted). }
\label{fig:dispar_with_theta}
\end{figure}

We also investigate how the disruption probability depends on the inclination angle 
$\theta$. As we have discussed in the previous section, since there is correspondence between 
$(\theta,\varphi)$ and $(\pi-\theta,\pi-\varphi)$, the disruption probability (and the 
energy averaged over binary phase) should be the same for the two binary orientations: 
$P_{dis}(\theta,\varphi)=P_{dis}(\pi-\theta, \pi-\varphi)$. By integrating this relation 
with respect to $\varphi$, we obtain the symmetry about $\theta=\pi/2$: 
$P_{dis}(\theta)=P_{dis}(\pi-\theta)$. The numerical disruption probability is shown 
in Figure \ref{fig:dispar_with_theta} as a function of $\theta$ for a fixed $D$.  We can clearly see such symmetry about $\theta=\pi/2$.

For deep penetrators $D\ll 1$, where the trajectory of the binary's centre of 
mass becomes radial, the binary orientation should be characterized only by the inclination 
angle $\theta$ (i.e. the angle between the radial direction and the binary rotation axis). 
Prograde or retrograde has no meaning or influence in this limit, indeed 
in Figure \ref{fig:dis_with_D}, 
prograde and retrograde results overlap in this regime. 
The radial approximation (the blue circles in the upper panel 
of Figure \ref{fig:dispar_with_theta}) 
is in a good agreement with the very deep penetrations ($D=10^{-5}$, the blue solid line) and the fractional difference in the probability of disruption between the radial and parabolic approximations becomes less than 2\% for $D<10^{-4}$.
Almost all binaries will be disrupted when the binary rotation axis is roughly oriented toward the BH or it is in the opposite direction. 
However, the surviving probability becomes significant for
$0.15\pi \lsim \theta \lsim 0.85\pi$, the highest surviving probability (or the lowest disruption probability $\sim 80\%$) is achieved for $\theta=\pi/2$. 
For larger values of $D$, the surviving probability increases for values of $\theta$ closer to $0$ and $\pi$. 

For very shallow penetrators, the highest disruption probability is archived around 
$\theta=\pi/2$, rather than $\theta\sim 0$ or $\pi$ (see the black dotted line in the 
bottom panel). This is because the dynamics depends on the relative orientation 
of the binary and orbital angular momenta for shallow penetrators, coplanar prograde orbits are relatively vulnerable to disruption.

\begin{figure}
\includegraphics[width=\columnwidth]{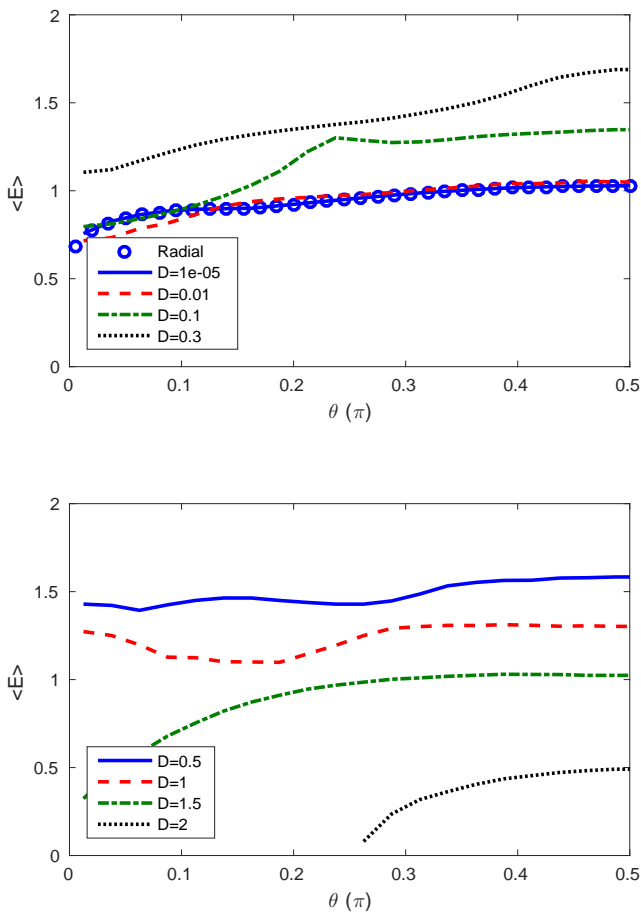}
\caption{Ejection energy averaged over binary phase and azimuth angel for a given $D$
as a function of the inclination angle $\theta$.
Radial solution (blue crosses) and parabolic solutions with $D=10^{-5}$ (blue solid line), 
$D=10^{-2}$ (red dashed-line), $D=0.1$ (green dot-dashed line), $D=0.3$ (black dotted line).
The average energy is in units of $(Gm_1 m_2 /a) (M/m)^{1/3}$, and it is 
evaluated for the absolute value of the energy $|E|$. }
\label{fig:meanE_with_theta}
\end{figure}

Figure \ref{fig:meanE_with_theta} indicates the ejection energy averaged over binary phase and the azimuth angle for a given $D$ as a function of the inclination angle $\theta$. As we have discussed,
the average energy is symmetric about $\theta=\pi/2$, the numerical results are plotted 
for $0< \theta < \pi/2$. The radial approximation results (the blue circles) and the 
parabolic approximation results for $D=10^{-5}$ (the blue solid line) are almost identical 
in this figure. However, there is a discrepancy at $\theta=0$. Due to the nature of the radial 
approximation binaries with $\theta=0$ have zero energy at all times. The parabolic approximation gives non-zero energy and its energy distribution is smooth around $\theta=0$.  For a wide range of $D$, 
the average energy slightly increases as the inclination angle $\theta$ increases. 
For the parabolic approximation results, the energy for $\theta=\pi/2$ is higher 
by a factor of $1.4-1.7$ than that for $\theta=0$.
Since the eccentricity differences $1-e$
of bound orbits are proportional to their orbital energy, the bound orbits are slightly less eccentric for $\theta =\pi/2$. 
However, as we have discussed, the eccentricities of the S-stars are determined by post-capture relaxation processes.

\subsection{Deep Encounter Survivors}

The existence of surviving binaries for $D\ll 1$ was first discussed
by our group \citep{2010ApJ...708..605S}. Recently \cite{2015arXiv150107856A} also reported a population of 
such surviving binaries in their large Monte Carlo simulations. 
Although deep encounter survivors are counter-intuitive, all binaries including these peculiar ones are actually disrupted when deeply penetrating the tidal sphere, and the binary members separate. However, they approach each other after the periapsis passage and a small fraction of them can form binaries again. 

To discuss this behaviour in more detail, we consider the radial restricted three-body 
approximation. Since the binary orientation is described by a single parameter in this regime, the discussion is simpler.  As we have discussed in section 2.1, we have analytic solutions eq. (\ref{freerad}) when the  binary deeply penetrates the tidal sphere. Since we have a set of three linear differential equations of the 2nd order, all solutions are linear combination of six independent solutions. Each could be physically obtained by taking the difference between an orbit infinitesimally close to a radial orbit and the radial orbit itself. In eq. (\ref{freerad}), the $A_x$ solution describes two particles that have the same trajectory, but are slightly separated in time. 
The $B_x$ solution describes the relative orbits of two particles going on the same radial path, but with slightly different energies. The energy gain or loss at the tidal encounter is proportional to $B_x$ (see \cite{2010ApJ...708..605S} for the full discussion). 

From eq. (\ref{freerad}), we can see that the $A_x$ solution dominates as the binary 
approach the "periapsis" ($t=0$ ; note that the radial approximation corresponds to the parabolic approximation in the deep penetration limit). In other words, the binary members are always in the same radial trajectory in the final approach stage, but they are separated in time. As the binary deeply penetrates the tidal sphere, the binary members separate wider and wider in the radial direction. However, they always approach each other after the periapsis passage  $x \sim A_x |t|^{-1/3}$. 

Although they approach each other, since the other free solutions begin to grow at $t>0$ (the index of the $B_x$ solution is the largest), they separate again in most cases. Only a small fraction of the pairs come out the tidal sphere as a binary. Although we do not fully understand the condition which ensures the binary formation after the periapsis passage, we have interesting results which indicate that the $A_x$ and $B_x$ solutions are likely to be related to the process.  

\begin{figure}
\includegraphics[width=\columnwidth]{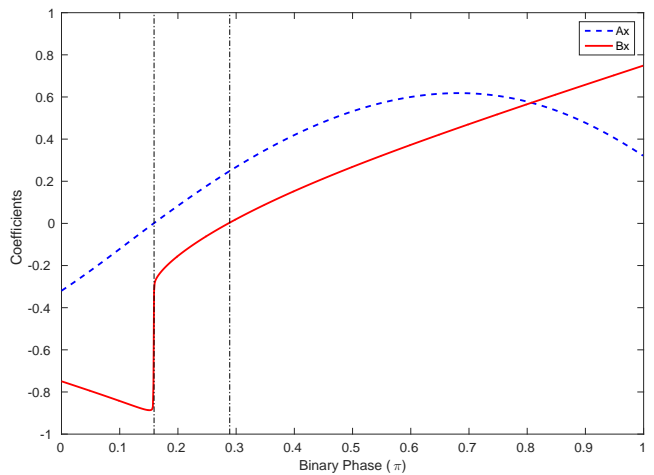}
\caption{the coefficients of the free solutions as functions of the initial binary phase:
 $A_x$ (blue dashed) and $B_x$ (red solid). 
the binary inclination $\theta=0.3\pi$.
Binaries survive the deep tidal encounter if the initial binary phase is in a narrow range indicated by the vertical black dot-dashed lines. }
\label{fig:AB1}
\end{figure}

Figure \ref{fig:AB1} shows the range of the initial binary phase $\phi$ for which binaries survive the deep encounter (i.e. binary formation after the periapsis passage). This is obtained based on the radial restricted three-body approximation with the binary inclination angle $\theta=0.3\pi$. The coefficients $A_x$ (the blue dashed line) and $B_x$ (the red solid line) are also shown as functions of $\phi$. These coefficients are evaluated at $t=-t_{min}=-10^{-6}$, and they are expected to become constants if the binary is disrupted (and the members separate widely). 
We notice interesting behaviours of the lines  at the boundaries of the surviving 
region (the vertical dot-dashed lines). $A_x$ becomes zero and the value of $B_x$ jumps at the left boundary, and $B_x$ is close to zero at the right boundary. 
The energy gain/loss of the binary members is proportional to $B_x$, large energy gain/loss near the surviving region has been reported in the previous study (e.g. Fig 6 in \cite{2010ApJ...708..605S}).  Although we have plotted the surviving range and the coefficients for $0<\phi<\pi$, 
a binary starting with a phase difference $\pi$ will have the same results (i.e. disrupted or not) and the same coefficients in absolute values but the opposite in sign. 

\begin{figure}
\includegraphics[width=\columnwidth]{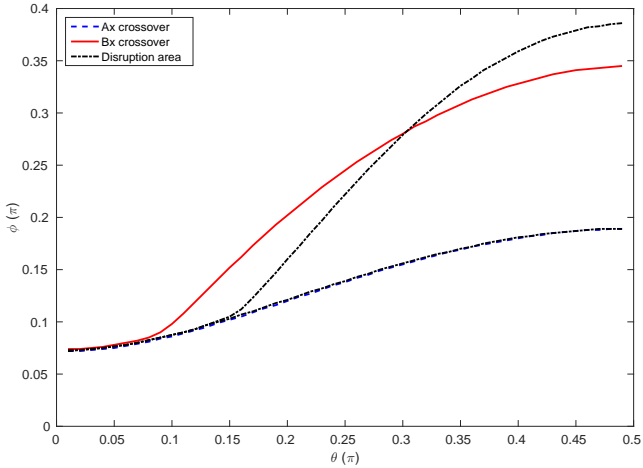}
\caption{The positions in the parameter space for which the coefficients of the free solution $A_x$ (blue dashed) and $B_x$ (red solid) are equal to zero and the boundaries of the range of binary phase for which binaries survive the deep tidal encounter (black dot-dashed).}
\label{fig:AB2}
\end{figure}

Figure \ref{fig:AB2} shows how the boundaries of the surviving region (the black dot-dashed lines) and the initial binary phases at which $A_x=0$ (the blue solid line) or $B_x=0$ (the red solid line) depend on the binary inclination angle $\theta$. At a large inclination angle (e.g. $\theta=\pi/2$), binaries survive the tidal encounter for a wide range of $\phi$. As a smaller inclination angle is assumed, the surviving region becomes narrower, and there are practically no survivors for $\theta < 0.15\pi$ (or $\theta > 0.85\pi$). 
In the figure, the $A_x=0$ line is identical to one of the boundaries of the surviving region (the lower branch). 
If $A_x$ is zero, the binary is just tidally compressed (i.e. no tidal stretch) when it approach the periapsis. 
Although the $B_x=0$ (or equivalently $E=0$) 
line is similar to the other boundary (the upper branch) of the surviving region, 
they are slightly different. We notice that the value of $B_x$ slightly evolve
even at $t>0$ around the boundary, because the binary members do not separate 
quickly in this region and they weakly interact each other. The real $E=0$ line 
is expected to be identical to the upper branch of the boundaries or slightly inside the surviving region. Otherwise, it means that some binaries are disrupted even if the energy gain or loss at the tidal encounter is zero ($E=0$). If the binary inclination angle $\theta$ is zero or $\pi$ (i.e. the binary rotation axis is exactly oriented toward the BH or it is in the opposite direction), 
from the symmetry of the system, the three-body interaction does not depend on the binary phase $\phi$ and we obtain $A_x=B_x=0$ for any $\phi$.

As we have just shown, for a given penetration depth and binary orientation, 
the fate of binaries (disrupted or not) is determined by the initial 
binary phase $\phi$. If $\phi$ is in a narrow surviving region, 
the binary survives the tidal encounter. It means that we can determine 
the probability of disruption accurately by resolving the narrow region 
with high resolution grid points. The advantage of our method is that 
we can analytically handle some of the system parameters
(e.g. masses of the binary members, initial binary separation, 
binary-to-BH mass ratio). The number of essential parameters is smaller 
than that for the full 3-body calculations. This allows us to set up 
high resolution grid points in the parameter space, rather than doing 
random sampling in the parameter space. 

We had checked the numerical convergence of our numerical results. For example, 
the probability of disruption shown in Figure \ref{fig:dis_with_D} (general orientations, the black solid line) is obtained with $N_{ori}=2000$ equally spaced grid points on a unit sphere (the orientation of a binary) and $N_{pha}=200$ equally spaced grid points for the binary phase. The results are about $87\%$ for $D=10^{-3}$ and $10^{-1}$. The probability is evaluated by changing $N_{ori}$ or $N_{pha}$ by a factor of $1/4-4$. The probability changes less than $0.3\%$ for the lower resolution (a factor of 1/4-1), and less than $0.05\%$ for the higher resolutions (a factor of 1-4). 

\subsection{Eccentric Binaries} 
We now consider the tidal disruption of coplanar, eccentric binaries. 
As we have discussed in the previous section, we sample $\varpi$ uniformly between 
0 and $\pi$. Since eccentric binaries spend a larger fraction of their time near the 
apoapsis, the binary phase $0<\phi< 2\pi$ is sampled with unequally 
spaced grids with which the binary rotates from a grid point to the next one with a constant 
time step. The time-averaged binary separation is given by $\bar{a}=a(1+e^2/2)$ 
where $a$ is the semi-major axis.

\begin{figure}
\includegraphics[width=\columnwidth]{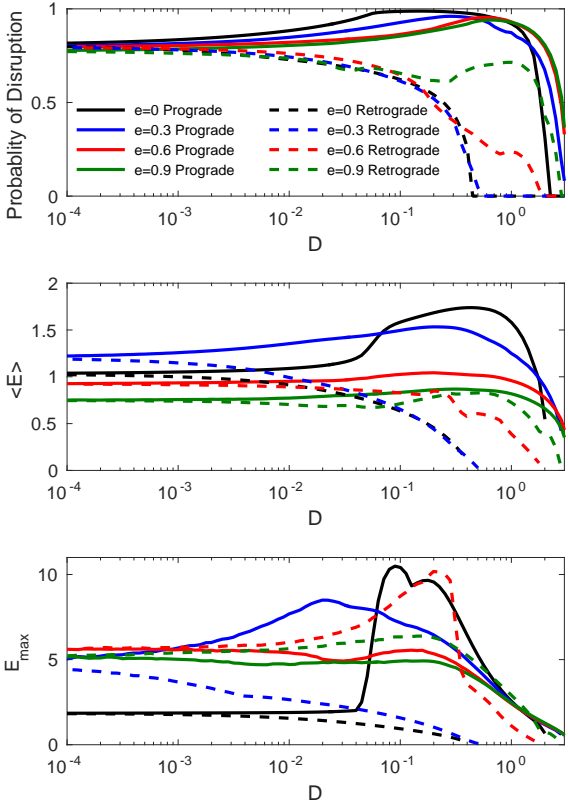}
\caption{Top panel: Probability of disruption. 
Middle panel: Ejection energy averaged over binary phase and the orientation of 
the semi-major axis.
Bottom panel: Characteristic  maximum ejection energy. These quantities are plotted 
as a function of the penetration factor $D$ for coplanar prograde (solid) and 
retrograde (dashed) orbits with $e = 0$, $0.3$, 
$0.6$, \& $0.9$ (black circles, blue crosses, red plusses, \& green triangles, respectively). 
The average and characteristic maximum energy 
are in units of $(Gm_1 m_2 /a) (M/m)^{1/3}$, and they are evaluated for the absolute value of 
the energy $|E|$.}
\label{fig:eccentric_dis_E}
\end{figure}

The top panel of Figure \ref{fig:eccentric_dis_E} shows the disruption probability of eccentric binaries 
as a function of $D$. All cases give $\sim 80\%$ disruption probability for $D\ll 1$. 
However, for shallow penetrators, the disruption probability strongly depends on 
the eccentricity and the direction of the binary rotation.
For prograde orbits (the solid lines), as higher eccentricity is assumed, the peak is shifted 
at a larger $D$, and the largest penetration factor $D_{max}$ 
for which there is disruption also becomes larger. $D_{max}$ is $\sim 2.1$ for $e=0$, 
$\sim 2.8$ for $e = 0.3$ and $\sim 3.2$ for $e = 0.6$ and $0.9$.
Since we have defined the penetration factor $D\propto a^{-1}$ by using the 
semi-major axis $a$, the effective binary separation $\bar{a}$ is larger than $a$, and 
consequently the effective penetration factor define with $\bar{a}$ is smaller by 
a factor of $(1+e^2/2)$ than $D$. For higher eccentricity, binaries are disrupted at 
a larger value of $D$, and the peak is shifted at a larger $D$. Although this qualitatively 
explains the shifts, the actually shits are larger (i.e. eccentric binaries are more 
vulnerable than circular ones at shallow encounter).
For retrograde orbits (the dashed lines), the eccentricity more significantly 
affects the probability distribution at shallow encounter.  Although the results for $e=0$ 
and 0.3 are similar, the probability distributions for $e=0.6$ and 0.9 have a peak 
structure around $D=1$, and  $D_{max}$ is much larger than the circular case ($D=0.44$ 
for $e=0$), they are comparable to the values for the corresponding prograde orbits. 

We plot the ejection energy averaged over the binary phase $\phi$ and orientation $\varpi$ 
in the middle panel of Figure \ref{fig:eccentric_dis_E}. We have scaled the energy by using the semi-major axis $a$
as $(Gm_1m_2/a)(M/m)^{1/3}$. 
For prograde orbits (the solid lines), as higher eccentricity is assumed, 
the distribution becomes flatter. The peak structure around $D=0.1-1$ which is 
significant for circular binaries ($e=0$; the black solid line) disappears. 
For retrograde orbits, for higher eccentricity, the distribution extends to 
larger $D$ because of a larger $D_{max}$, and the distribution becomes similar 
to the prograde one.
In the deep penetration limit $D\ll 1$, the prograde and retrograde orbit cases 
approach the same ejection energy as expected.  Interestingly, the asymptotic energy 
is not a monotonic function of the eccentricity,  the largest value is given by 
$e=0.3$ (the blue lines).

The distributions of the characteristic maximum energy $E_{max}$ also behave in a 
similar way especially for the prograde orbits (the solid lines in 
the bottom panel): the distributions becomes flatter for higher eccentricity. 
However, the distribution for retrograde orbits (the dashed lines) have a peak 
structure for high eccentricity, it is significant especially for $e=0.6$ (the red dashed line).
The values around $D=0.1$ become even larger than the corresponding prograde cases 
for $e=0.6$ and 0.9 (the red and green lines).
The asymptotic values at $D\ll1$ are similar for prograde and retrograde orbits, and 
higher eccentricity gives a higher value.

\begin{figure}
\includegraphics[width=\columnwidth]{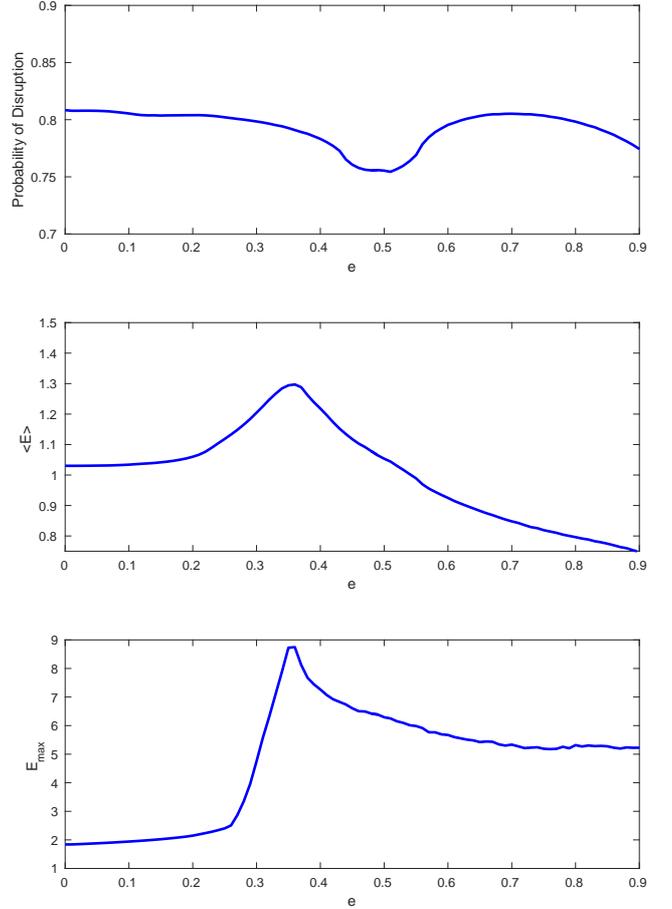}
\caption{Radial approximation results for the inclination angle $\theta=\pi/2$.
Top panel: Probability of disruption as a function of eccentricity.
Middle panel: Ejection energy averaged over binary phase and the direction of the 
semi-major axis as a function of eccentricity. 
Bottom panel: Characteristic maximum ejection energy as a function of eccentricity.
The average and characteristic maximum energy are in units of $(Gm_1 m_2 /a) (M/m)^{1/3}$, 
and they are evaluated for the absolute value of the energy $|E|$.}
\label{fig:radial_ecc}
\end{figure}

The asymptotic values at $D\ll 1$ are also estimated by using the radial approximation. 
Although the disruption probability at $D\ll 1$ is less sensitive to the eccentricity
(see the top panel of Figure \ref{fig:radial_ecc}), there is a small dip around $e=0.5$. 
The average ejection energy has been scaled by $(Gm_1m_2/a)(M/m)^{1/3}$. 
Since the effective binary separation $\bar{a}$ is larger than the semi-major axis, 
the disruption of a wider binary should result in an ejection energy smaller 
by a factor of $(1+e^2/2)$. 
The numerical results show a smaller energy for $e=0.9$, compared to the circular 
case $e=0$ (see the middle panel of Figure \ref{fig:radial_ecc}), and the number is roughly consistent:
$1/(1+e^2/2)\sim 0.7$. However, 
the numerical energy peaks around $e= 0.35$. The eccentricity more drastically 
affects $E_{max}$ (see the bottom panel). The values at $e > 0.4$ is much larger than 
that for the circular case, and there is a significant peak around $e= 0.35$.

\section{Conclusions and Discussion}

We have discussed how binary tidal disruption depends on the inclination and 
eccentricity of the binary. When the binary-to-BH mass ratio is large $M/m \gg 1$, 
our restricted three-body approximation allows us to explore the parameter space efficiently.
For inclined, circular binaries, we have show that about $12\%$ of them with random 
orientations survive even if they approach the massive BH very closely: $D\ll 1$.
Although the existence of the surviving binaries is counter-intuitive, 
the binary members actually once separate even in the surviving cases, 
and approach each other again after their periapsis passage. This surviving probability 
is lower than $\sim 20\%$ obtained for coplanar cases \citep{2010ApJ...708..605S}. This is because almost 
all deep penetrators are disrupted when the binary rotation axis is 
roughly oriented toward the massive BH or in the opposite direction 
(i.e. the inclination $\theta \lsim 0.15\pi$ or $\theta \gsim 0.85\pi$ ).
The maximal surviving probability is achieved for $\theta = \pi/2$ for a wide range of 
$D$, however, if $D$ is close to the largest $D$ for which there is disruption,
disruption is only found in prograde co-planar orientations.
The average energy $\langle E \rangle$ also depends on the inclination $\theta$, but 
the dependence is weak. The energy for $\theta=\pi/2$ is higher 
by a factor of $1.4-1.7$ than that for $\theta=0$. 

Our coplanar calculations show that the disruption probability at $D\ll 1$ is 
insensitive to the eccentricity of binaries, all cases of prograde and retrograde 
orbits with $e=0-0.9$ give very similar disruption probability.
The ejection energy at $D\ll 1$ is more sensitive to the eccentricity. 
This can be partially explained by an effectively wider binary 
separation for more eccentric binaries. However, the energy is not a monotonic function 
of the eccentricity, and it peaks at $e\sim 0.35$.  For shallow penetrators,
both disruption probability and ejection energy are more strongly affected by the 
eccentricity, especially in retrograde orbits where disruption
rates become closer to that of prograde orbits with higher eccentricity. 

Our results were obtained assuming point-like stars. For $D \ll1$, the pericentre distance to the central MBH becomes comparable to the tidal radius of the individual stars that compose the binary system.  
If the binary members are solar-type stars 
with its initial separation $a \sim 1$ AU, they are tidally disrupted 
for $D\lsim 5\times10^{-3}$.
Such double tidal disruption events have been discussed by \cite{2015ApJ...805L...4M}. To achieve a deeper penetration without the disruption of the binary members, binaries need to have a wide initial separation, or they should consist of compact objects. We intend to provide the basic characteristics of the tidal encounter between binaries and a massive object in this paper.
The tidal disruption of stellar mass 
BH binaries will be investigated in the context of BH mergers and LIGO observations (Fernandez et al. in preparation). Another possible implication of our results is the study of irregular satellites around giant planets, 
they follow a distant, inclined, and often eccentric and retrograde orbit.
One of the leading mechanisms to produce such satellites is the three-body tidal encounter
\citep{2006Natur.441..192A,2012ApJ...748..105K}. 

We do not account for the possibility of stellar collisions 
during the tidal encounter. However, such collisions
and the resultant mergers could have interesting consequences
\citep{2017MNRAS.469.2042B}. We roughly evaluate the collision rate by using the parabolic 
restricted three-body approximation. Although the energy and disruption probability are 
accurately evaluated in this approximation, the separations of the binary members are 
overestimated for a short period around the peripasis passage $|t| < 
(m/M)^{1/3} \sqrt{a^3/Gm}$ for $D< (m/M)^{1/3}$ \citep{2010ApJ...708..605S}. 
If the mass ratio $M/m$ is not very large, we might underestimate the collision rate. For $a/R_\star=10$ where $a$ is 
the initial binary separation and $R_\star$ is the radius of the binary members,
we have evaluated the collision probability at the BH encounter as a 
function of $D$ averaged over phase and orientation. If the minimum separation of 
binary members becomes less than $2R_\star$ during the evolution (or equivalently 
if it becomes less than $1/5$ of the initial separation), we regard it as a 
collision case. 
We find that the collision probability is about 5-7 $\%$ for $D<0.1$, 
and that the probability slightly increases at shallow encounters 
and it peaks around $D=1.6$ with a peak value of $\sim 14\%$.
Even if collisions are taken into account, the disruption probability is very similar. 
Compared to the point particle results, the fractional difference $\Delta P_{dis}/P_{dis}$ is a few $\%$ for $D<0.1$, and it peaks around $D=1$ with $\Delta P_{dis}/P_{dis} \sim 5\%$.
It means that most collision events are classified into the surviving case in 
the point particle calculations. $\Delta P_{survive}/P$ is about $45\%$ for 
$D<3\times 10^{-2}$ (i.e. for deep penetrators, the surviving probability becomes about 
a half of the point particle value). $\Delta P_{survive}/P_{survive}$ 
gradually decreases for larger $D$ and it is about 20$\%$ for $D\sim 1$.

\section*{Acknowledgments}
We thank the anonymous referee for valuable suggestions. 
This research was supported by STFC grants.

\bsp
\label{lastpage}

\begin{thebibliography}{999}
\bibitem[Addison et al.(2015)]{2015arXiv150107856A} Addison, E., Laguna, P., \& Larson, S.\ 2015, arXiv:1501.07856
\bibitem[Agnor \& Hamilton (2006)]{2006Natur.441..192A} Agnor,C.B., \& Hamilton, D.P. \ 2006, Nature, 441, 192
\bibitem[Alexander (2017)]{2017ARAA.55.1} Alexander, T. \ 2017, Annu. Rev. Astron. Astrophys. 55,1
\bibitem[Antonini et al.(2010)]{2010ApJ...713...90A} Antonini, F., Faber, J., Gualandris, A., \& Merritt, D.\ 2010, \apj, 713, 90 
\bibitem[Bradnick et al.(2017)]{2017MNRAS.469.2042B} Bradnick, B., Mandel, I., \& Levin, Y.\ 2017, \mnras, 469, 2042
\bibitem[Bromley et al.(2006)]{2006ApJ...653.1194B} Bromley, B.~C., Kenyon, S.~J., Geller, M.~J., et al.\ 2006, \apj, 653, 1194 
\bibitem[Bromley et al.(2012)]{2012ApJ...749L..42B} Bromley, B.~C., Kenyon, S.~J., Geller, M.~J., \& Brown, W.~R.\ 2012, \apjl, 749, L42 
\bibitem[Brown et al.(2009)]{2009ApJ...690.1639B} Brown, W.~R., Geller, M.~J., \& Kenyon, S.~J.\ 2009, \apj, 690, 1639 
\bibitem[Brown et al.(2012)]{2012ApJ...751...55B} Brown, W.~R., Geller, M.~J., \& Kenyon, S.~J.\ 2012, \apj, 751, 55 
\bibitem[Brown et al.(2014)]{2014ApJ...787...89B} Brown, W.~R., Geller, M.~J., \& Kenyon, S.~J.\ 2014, \apj, 787, 89 
\bibitem[Ghez et al.(2008)]{2008ApJ...689.1044G} Ghez, A.~M., Salim, S., Weinberg, N.~N., et al.\ 2008, \apj, 689, 1044-1062 
\bibitem[Gillessen et al.(2009)]{2009ApJ...692.1075G} Gillessen, S., Eisenhauer, F., Trippe, S., et al.\ 2009, \apj, 692, 1075 
\bibitem[Ginsburg \& Loeb(2006)]{2006MNRAS.368..221G} Ginsburg, I., \& Loeb, A.\ 2006, \mnras, 368, 221
\bibitem[Gualandris et al.(2005)]{2005MNRAS.363..223G} Gualandris, A., Portegies Zwart, S., \& Sipior, M.~S.\ 2005, \mnras, 363, 223 
\bibitem[Hills(1988)]{1988Natur.331..687H} Hills, J.~G.\ 1988, \nat, 331, 687 
\bibitem[Hills(1991)]{1991AJ....102..704H} Hills, J.~G.\ 1991, \aj, 102, 704.
\bibitem[Kobayashi et al.(2012)]{2012ApJ...748..105K} Kobayashi, S., Hainick, Y., Sari, R., \& Rossi, E.~M.\ 2012, \apj, 748, 105 
\bibitem[Lu et al.(2010)]{2010ApJ...709.1356L} Lu, Y., Zhang, F., \& Yu, Q.\ 2010, \apj, 709, 1356 
\bibitem[Mandel \& Levin(2015)]{2015ApJ...805L...4M} Mandel, I., \& Levin, Y.\ 2015, \apjl, 805, L4 
\bibitem[Meibom \& Mathieu(2005)]{2005ApJ...620..970M} Meibom, S., \& Mathieu, R.~D.\ 2005, \apj, 620, 970 
\bibitem[Miller et al. (2005)]{2005ApJ...631..L117} Miller,M.C. Freitag, M. Hamilton,D.P. \& Lauburg, V.M. \ 2005, \apj, 631, L117
\bibitem[Perets et al.(2009)]{2009ApJ...702..884P}
Perets, H., Gualandris,A., Kupi, G. et al. \ 2009, \apj, 702, 884
\bibitem[Perets et al.(2007)]{2007ApJ...656..709}
Perets, H., Hopman,C. \& Alexander,T. \ 2007, \apj,656, 709
\bibitem[Pfahl(2005)]{2005ApJ...626..849P}
Phahl,E. \ 2005, \apj,626, 849
\bibitem[Rossi et al.(2014)]{2014ApJ...795..125R} Rossi, E.~M., Kobayashi, S., \& Sari, R.\ 2014, \apj, 795, 125 
\bibitem[Rossi et al.(2017)]{2017MNRAS.467.1844R} Rossi, E.~M., Marchetti, T., Cacciato, M., Kuiack, M., \& Sari, R.\ 2017, \mnras, 467, 1844 
\bibitem[Sari et al.(2010)]{2010ApJ...708..605S} Sari, R., Kobayashi, S., \& Rossi, E.~M.\ 2010, \apj, 708, 605 
\bibitem[Sesana et al.(2007)]{2007MNRAS.379L..45S} Sesana, A., Haardt, F., \& Madau, P.\ 2007, \mnras, 379, L45 
\bibitem[Yu \& Tremaine(2003)]{2003ApJ...599.1129Y} Yu, Q., \& Tremaine, S.\ 2003, \apj, 599, 1129 
\end{thebibliography}
\end{document}